\documentclass[12pt,a4paper]{article}
\usepackage{amsmath}
\usepackage[font={footnotesize,it}]{caption}
\usepackage[top=1.21in, bottom=1.21in, left=1.2in, right=1.2in]{geometry}

\DeclareCaptionStyle{italic}[justification=centering]{labelfont={bf},textfont={it},labelsep=colon}
\usepackage{graphicx,psfrag,epsf}
\usepackage{enumerate}
\usepackage{natbib}
\usepackage{url, setspace}
\usepackage{comment}
\usepackage{multirow}
\usepackage{subcaption}
\usepackage{booktabs, bm, paralist,mathpazo,tikz,longtable,microtype,dsfont,rotating} 
\usepackage[pdftex,colorlinks=true]{hyperref}
\definecolor{darkblue}{rgb}{0,0,.6}
\hypersetup{citecolor=darkblue,linkcolor=darkblue,urlcolor=darkblue}

\newcommand{\blind}{0}

\graphicspath{{plots/}}

\newsavebox\CBox

\date{}

\begin{document}

\def\spacingset#1{\renewcommand{\baselinestretch}
{#1}\small\normalsize} \spacingset{1}

\if0\blind
{
  \title{\bf  Change point detection for COVID-19 excess deaths in Belgium}
  \author{Han Lin Shang\thanks{Corresponding address: Department of Actuarial Studies and Business Analytics, Level 7, 4 Eastern Road, Macquarie University, NSW 2109, Sydney, Australia; Email: hanlin.shang@mq.edu.au; ORCID: \url{https://orcid.org/0000-0003-1769-6430}}\\
Department of Actuarial Studies and Business Analytics \\
 Macquarie University \\
\\
   Ruofan Xu
  \hspace{.2cm}\\
    Research School of Finance, Actuarial Studies and Statistics \\
    Australian National University}
  \maketitle
} \fi

\if1\blind
{
   \title{\bf Change point detection for COVID-19 excess deaths in Belgium}
   \author{}
   \maketitle
} \fi

\bigskip

\begin{abstract}
Emerging at the end of 2019, COVID-19 has become a public health threat to people worldwide. Apart from the deaths who tested positive for COVID-19, many others have died from causes indirectly related to COVID-19. Therefore, the COVID-19 confirmed deaths underestimate the influence of the pandemic on the society; instead, the measure of `excess deaths' is a more objective and comparable way to assess the scale of the epidemic and formulate lessons. One common practical issue in analyzing the impact of COVID-19 is to determine the `pre-COVID-19' period and the `post-COVID-19' period. We apply a change point detection method to identify any change points using the excess deaths in Belgium.
\vspace{.1in}

\end{abstract}

\newpage
\spacingset{1.58}

\section{Introduction}\label{S1}

Emerging at the end of 2019, COVID-19 has become a public health threat to people worldwide. According to statistics compiled by Johns Hopkins University \citep{Covid19DB}, up until early June, the COVID-19 pandemic has claimed nearly 400,000 lives globally. With a population of just 11.5 million, Belgium had the highest death to population ratio, with 78 deaths per 100,000 people \citep{CovidMortality} at that time. This figure includes only deaths directly linked to COVID-19 and is likely inaccurate due to different reporting practices or people dying of COVID-19 related causes without being tested. The physical, psychological, and social effects of lockdown and economic changes stemming from COVID-19 have indirectly caused deaths, which are not counted in official figures. Therefore, the deaths directly linked to COVID-19 underestimate the impact of the pandemic on society.

One way to explore the pandemic's actual mortality effect is to look at the number of `excess deaths'. This measure can be constructed by comparing the observed weekly deaths throughout 2020 to values from the previous non-pandemic period. This provides a more objective and comparable way to assess the scale of the pandemic and formulate lessons. 

Attempts have been made to continuously track and examine comparative excess mortality data for Europe, the United Kingdom, the United States of America, and other countries. Simple visualizations and discussions can be found in \cite{FT2020}, \cite{Eoconomist2020} and etc. Researchers have recently developed different approaches to understand the dynamics of the pandemic. In the field of demography, researchers have focused on the impact of COVID-19 on different demographic groups, such as age, gender, and income level. For example, \cite{dowd2020demographic} examine the role of age structure in deaths, showing that countries with a much older population have a dramatically higher mortality rate than those with a much younger population. \cite{cairns2020impact} analyzed the impact of COVID-19 on future higher-age mortality. \cite{riley2020estimates} assessed the potential impact of the COVID-19 pandemic on sexual and reproductive health in low-and middle-income countries. \cite{banerjee2020estimating} estimated the one-year excess mortality associated with the COVID-19 pandemic.

One common practical issue in analyzing the impact of COVID-19 is to classify the 'pre-COVID-19' period and the 'post-COVID-19' period. Some researchers choose the date when the first death was reported \citep[see, e.g.,][]{vandoros2020excess}, while the others choose the date when the number of fatalities began increasing steeply as the change point \citep[see, e.g.,][]{barberostatistical}. The former choice may underestimate the magnitude of any effect on non-COVID-19 death, while the latter choice is likely to exaggerate any findings. Hence, it is necessary to propose a statistical method for determining the breakpoint.

As age group \citep{dowd2020demographic} highly influences the COVID-19 mortality, a natural question to ask is whether the phase changes also exhibit age patterns? If this is the case, practitioners should classify the 'pre-COVID-19' and post-COVID-19' period for each age group independently when analyzing COVID-19 impact differences between age groups. In addition, the phases change differences among age groups may aid in establishing more targeted policy and public guidance towards different age groups in different pandemic periods.

To this end, a change point detection method for time series may provide a statistically reliable result for phase change detection of excess deaths during the COVID-19 pandemic. In the literature, change point analysis either deals with the mean change \citep{bai2003computation}, variance change \citep{hawkins2005change}, or the distributional change \citep{matteson2014nonparametric} within time-ordered observations. The number of change points is usually assumed to be single or a known number. However, as the COVID-19 pandemic may change the mean or variance of the underlying data structure, we seek to detect distributional changes. Using the hierarchical divisive estimation approach proposed by \cite{matteson2014nonparametric}, any distributional changes within an independent time-ordered sequence can be identified simultaneously without assuming the exact number of change points.

This paper's contribution is to provide a statistical change point detection method to identify the change points in the counts of COVID-19 caused excess death. The rest of the paper is organized as follows. In Section~\ref{s2}, we introduce the dataset and provide a brief analysis of the excess deaths in Belgium. The change point detection method is provided in Section~\ref{s3}, followed by a discussion of the change point detection results for the excess deaths in Belgium in Section~\ref{s4}. We examine if the phase changes exhibit age differences — finally, we summarize our search findings in Section~\ref{s5}.

\section{Methodology}

\subsection{Dataset overview}\label{s2}

To examine the excess death counts in Belgium, we extract the data from the~\cite{HMD}. The original data contains the weekly death counts in Belgium from January 1\textsuperscript{st}, 2010 to May 16\textsuperscript{th}, 2020. To obtain a stationary time series, we only use the past five-year data. In summary, we use the weekly data from week 1 in 2015 to week 20 in 2020, which are 279 weeks in total. The data are recorded by gender (male and female) and age group (age 0 - 14, 15 - 64, 65 - 74, 75 - 84, and above (and including) 85).

\begin{figure}[!htbp]
    \centering
    \includegraphics[width=0.8\linewidth]{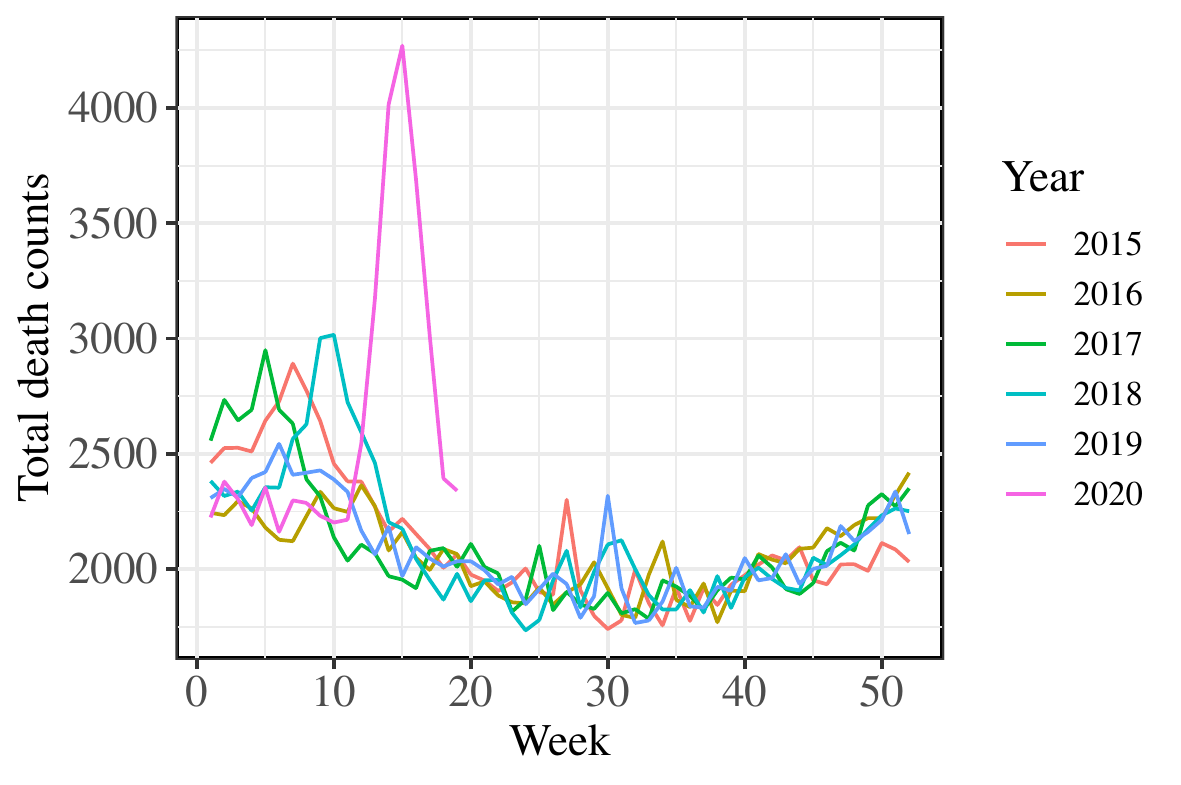}
    \caption{The weekly total death counts in Belgium}
    \label{data}
\end{figure}

Figure~\ref{data} presents the weekly total death counts from 2015 to the present plotted against the week of the year. There is a clear annual seasonal pattern due to temperature and other weather-related effects \citep[see also][]{healy2003excess}. The graph shows the COVID-19 pandemic effect in 2020, and it also indicates increased mortality rates at the start of 2015 and 2018 and the first half of March 2018. This increase is probably due to the flu epidemic.

To minimize the impact of large random fluctuation, we use the weekly median death counts in the past five years to capture the pattern of `pre-COVID-19' death counts. The excess death counts are computed as the difference between the actual weekly death counts and the weekly median death counts \citep{owidcoronavirus}. Then, to adjust for the subpopulation size, the excess death rate is computed as the excess death counts divided by the median death counts from 2015 to 2019. Compared with the excess death counts, the rate is more suitable for comparing the impact across groups. Therefore, in the following analysis, we work with the excess death rate labeled by gender and age groups from 2015 to the present.

\begin{figure}[!htbp]
    \centering
    \includegraphics[width=0.75\linewidth]{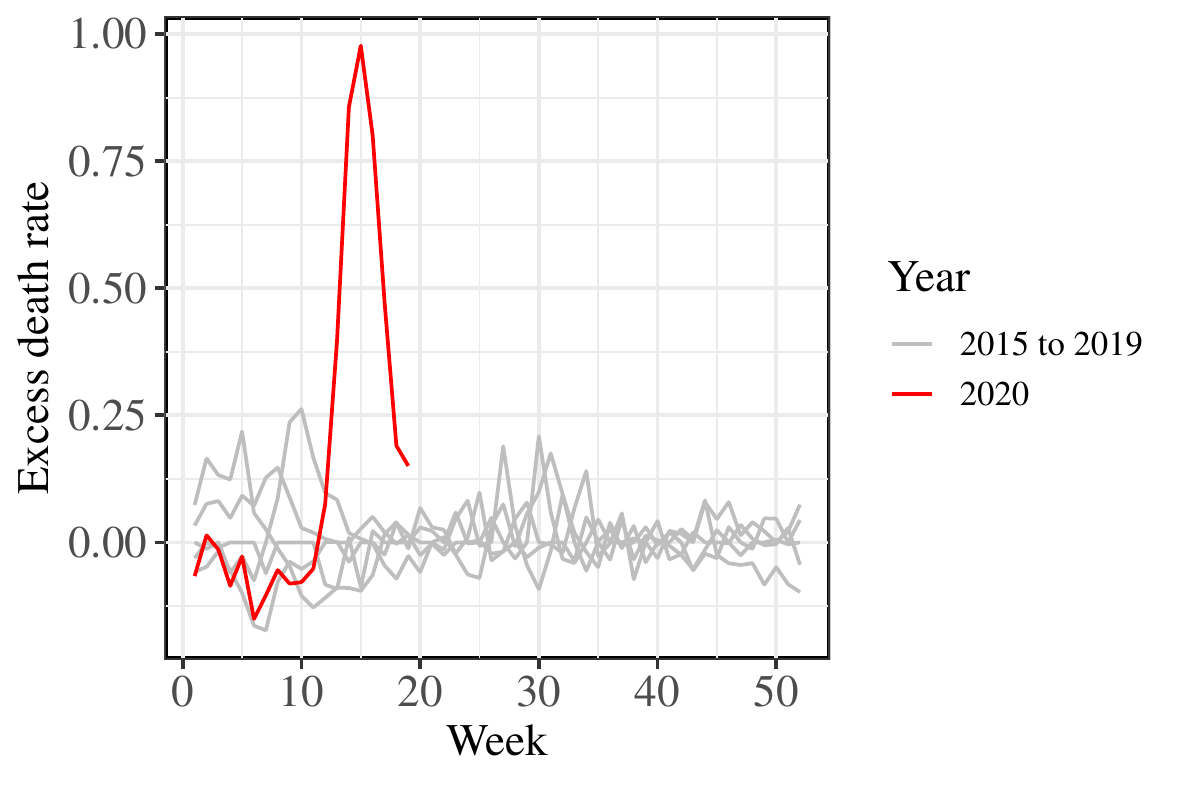}
    \caption{The weekly total excess death rate in Belgium}
    \label{excess_death}
\end{figure}

The weekly total excess death rates are shown in Figure~\ref{excess_death}, with the rates in 2020 highlighted in red. In early 2020 (before week 11), the weekly excess death rate is negative. We might not draw any conclusions based on this information as the deaths generally vary much at the beginning of a year. However, from week 13 (end of March), the weekly excess death rate increased dramatically, and it continued to increase until week 15. Based on the graphical interpretation, we have gained insight into the critical time points in the COVID-19 pandemic. In the following sections, we utilize statistical methods to identify the change points that mark the different phases in the COVID-19 pandemic.

\begin{figure}[!htbp]
    \centering
    \includegraphics[width=1\linewidth]{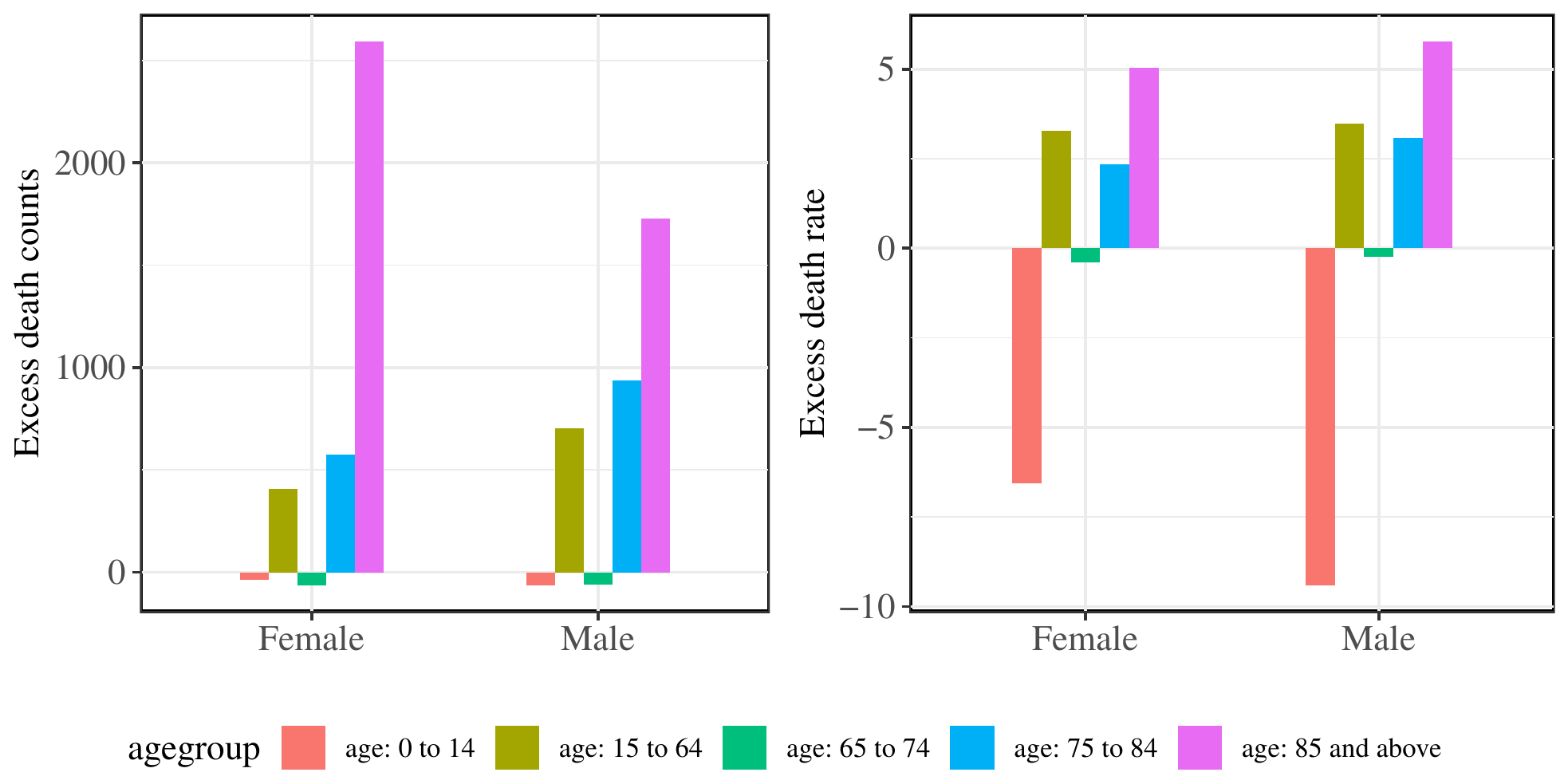}
    \caption{The total excess death counts and rate in 2020, grouped by age, for female and male subpopulations in Belgium}
    \label{excess_death_bar}
\end{figure}

The bar charts in Figure~\ref{excess_death_bar} show a different level of the negative impact of COVID-19 for different subpopulations in Belgium. Both genders show similar excess death patterns. Surprisingly, the total excess death rate in 2020 for ages between 65 and 74 is negative. However, when examining excess deaths by the week in Section~\ref{s4}, we notice that the excess deaths for thosed aged between 65 and 74 after week 13 are positive, which suggests that the elderly (age above 64) are more susceptible to COVID-19. In particular, those aged above 85 are more vulnerable to COVID-19 than other age groups. Meanwhile, COVID-19 has little impact on excess deaths amongst children and teenagers, as they travel and going out less. By contrast, COVID-19 is a much greater threat to the older people, as they generally suffer from the physiological changes associated with aging, decreased immune function, and multimorbidity. Additionally, the gender difference in excess death counts has different patterns among age groups. For age groups below 85, males generally have a higher excess death counts, while there are more excess deaths for females than males for those aged over 85. This is because the female subpopulation aged above 85 is larger than that of males. The median death count for females above 85 is also higher than males. Therefore, after we control for median death counts, males have higher death rates than females for all age groups. This phenomenon is consistent with the findings in many existing empirical studies that male tends to have more severe infection outcome (e.g., higher mortality rate) in all age groups \citep[see, e.g.,][]{haitao2020covid,bhopal2020sex}.

\subsection{Change point detection} \label{s3}

We consider a hierarchical divisive estimation approach \citep{szekely2005hierarchical} to detect the single or multiple change points in the excess death counts. We adopt this approach because it can detect a distribution change without assuming any specific distribution structure. Therefore, it is more flexible than \cite{bai2003computation} and \cite{hawkins2005change}, where only the mean and variance change can be detected.

Let $\{Z_t,t=1,\dots,T\}$ be an independent sequence of time-ordered observations (either univariate or multivariate). Given a change point $t^{*}$, the observed data is partitioned into two clusters $A_{t^{*}} = \{Z_{1},\dots, Z_{t^*}\}$ and $B_{T-t^{*}} = \{Z_{t^* + 1},\dots,Z_{T}\}$. Both $A_{t^*}$ and $B_{t^*}$ are independent and identically distributed (i.i.d) samples from the distribution $A,B \in \mathbb{R}^d$ for $d \in \mathbb{N}$, respectively, such that $\mathbb{E}|A|^{\alpha}, \mathbb{E}|B|^{\alpha}< \infty$ for some $\alpha \in (0,2)$. We compute the generalized energy distance between $A$ and $B$ as follows:
\begin{align}
\label{dist}
\widehat{\mathcal{E}}(A_{t^{*}},B_{T-t^{*}};\alpha)
& = \frac{2}{t^{*}(T-t^{*})}\sum_{i=1}^{t^*}\sum_{j=t^*+1}^{T}|Z_i - Z_j|^{\alpha} \\
& \quad - \frac{1}{\left(\begin{smallmatrix}
t^*\\
2
\end{smallmatrix}\right)} \sum_{i=2}^{t^*}\sum_{j=1}^{i-1}|Z_i - Z_j|^{\alpha}
- \frac{1}{\left(\begin{smallmatrix}
T-t^*\\
2
\end{smallmatrix}\right)} \sum_{i=t^*+2}^{T}\sum_{j=t^*+1}^{i-1}|Z_i - Z_j|^{\alpha}, \nonumber
\end{align}
where $|\cdot|$ is the Euclidean norm.

The first term on the right-hand side of~(\ref{dist}) corresponds to the between-distance measure of $A_{t^*}$ and $B_{T-t^*}$. The second and third terms on the right-hand side of~(\ref{dist}) correspond to the within-distance of $A_{t^*}$ and $B_{T-t^*}$, respectively. To incorporate different sample sizes, \cite{szekely2005hierarchical} proposed a $Q$ statistic defined as 
\begin{align*}
\widehat{Q}(A_{t^*},B_{T-t^*};\alpha) = \frac{t^* (T-t^*)}{T}\widehat{\mathcal{E}}(A_{t^{*}},B_{T-t^{*}};\alpha).
\end{align*}
A change point location $t^*$ is determined as the value that maximizes $\widehat{Q}(A_{t^*},B_{T-t^*};\alpha)$. For detecting a difference in mean, $\alpha$ is set to 2. 

To estimate multiple breakpoints, we iteratively apply the above techniques as follows. Suppose that $k-1$ change points have been estimated at time $0=\widehat{t}_0 < \widehat{t}_1 < \dots < \widehat{t}_{k-1} < \widehat{t}_k = T$. These change points can partition the observations into $k$ clusters ($\widehat{C}_1,\dots,\widehat{C}_k$), such that $\widehat{C}_i = \{Z_{\widehat{t}_{i-1}+1},\dots,Z_{\widehat{t}_{i+1}}\}$. Given these clusters, we then apply the procedure for finding a single change point to the observations within each of the $k$ clusters, so that we obtain $k$ candidate change points. Then we choose the one resulting in the largest $Q$ statistics as the estimated $k^{\text{th}}$ change points. Then, we check if the selected $k^{\text{th}}$ change point is significant using a permuatation hypothesis test. Details of the hypothesis test can be found in \cite{szekely2005hierarchical}. The iteration terminates if the $k^{\text{th}}$ estimated change point tested is insignificant. The implementation of this distance-based approach is included in the $\texttt{ecp}$ package \citep{james2014ecp} in R \citep{Team20}.

\section{Results and discussion} \label{s4}

Since our interest lies in the change point detection caused by COVID-19, we perform the change point detection on data from the latter half of 2019 until the most recent observation in 2020. 

Let $\{\mathbf{X}_t,t=1,\dots,T\}$ be the observed excess death rate starting from the 27\textsuperscript{th} week of Year 2019, so that $T=46$ and $\mathbf{X}_t = \left(X^{(f)}_{t,i},X^{(m)}_{t,j},i,j=1,\dots,5\right)^{\top}$. The superscript `\emph{m}' and `\emph{f}' are used to distinguish the gender `male' and `female', and the subscript $t$ denotes the week index and $i,j$ denotes the five age groups (age 0 - 14, 15 - 64, 65 - 74, 75 - 84 and above (and including) 85). 

\begin{figure}[!htbp]
    \centering 
  \begin{subfigure}[b]{0.33\textwidth}
                \caption{Age between 0 and 14}
                \includegraphics[width=\linewidth]{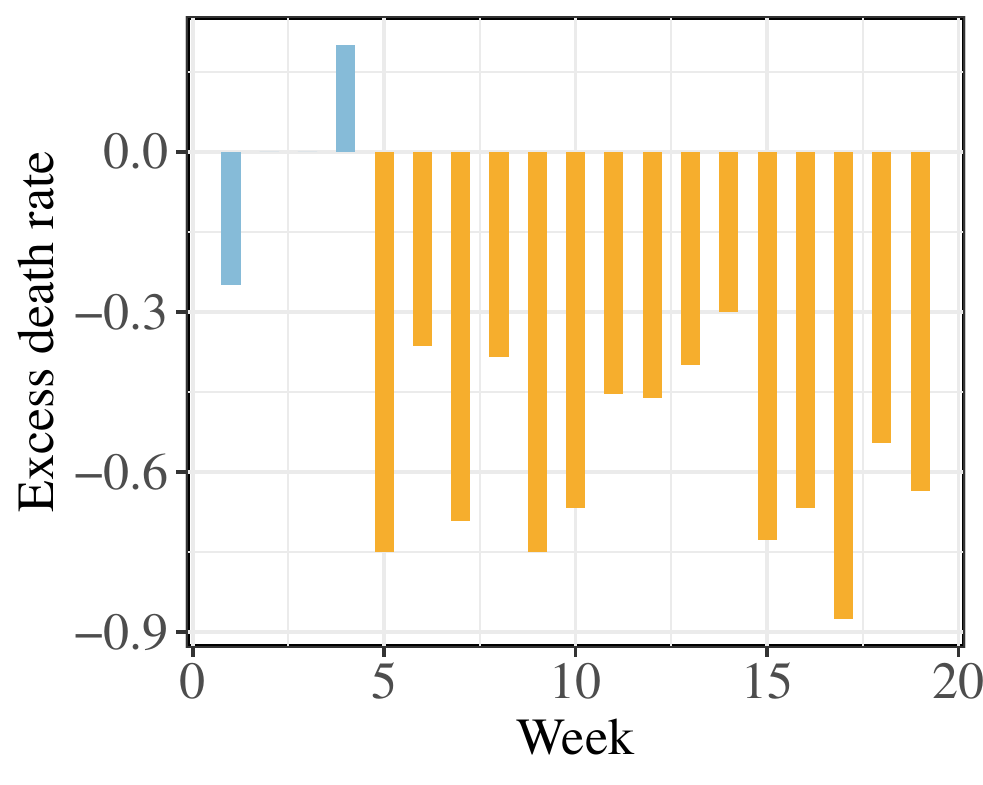}
                \label{p1}
    \end{subfigure}%
    \begin{subfigure}[b]{0.33\textwidth}
                \caption{Age between 15 and 64}
                \includegraphics[width=\linewidth]{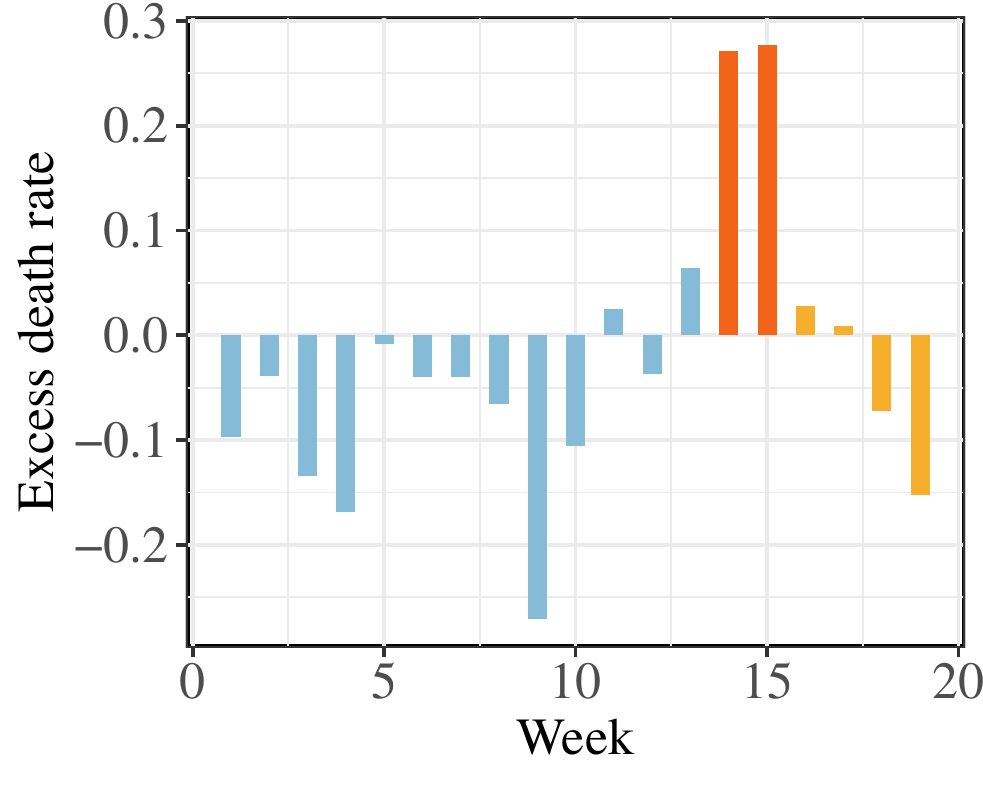}
                \label{p2}
    \end{subfigure}
    \begin{subfigure}[b]{0.33\textwidth}
                \caption{Age between 65 and 74}
                \includegraphics[width=\linewidth]{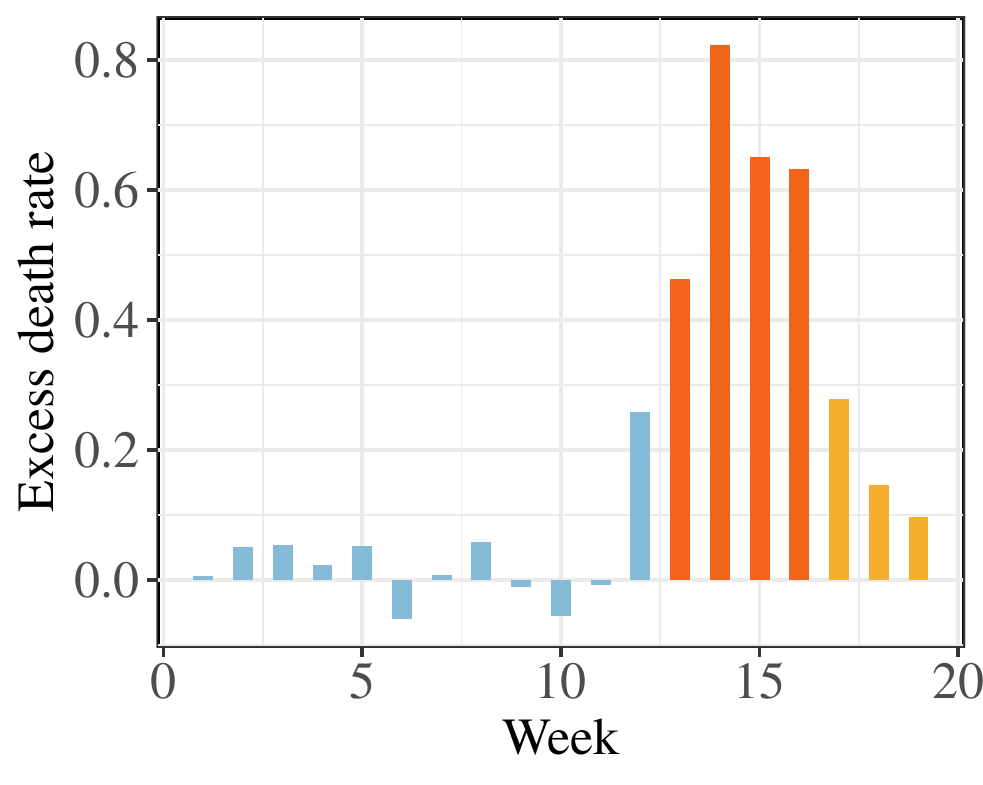}
                \label{p3}
    \end{subfigure}
    \begin{subfigure}[b]{0.33\textwidth}
                \caption{Age between 75 and 84}
                \includegraphics[width=\linewidth]{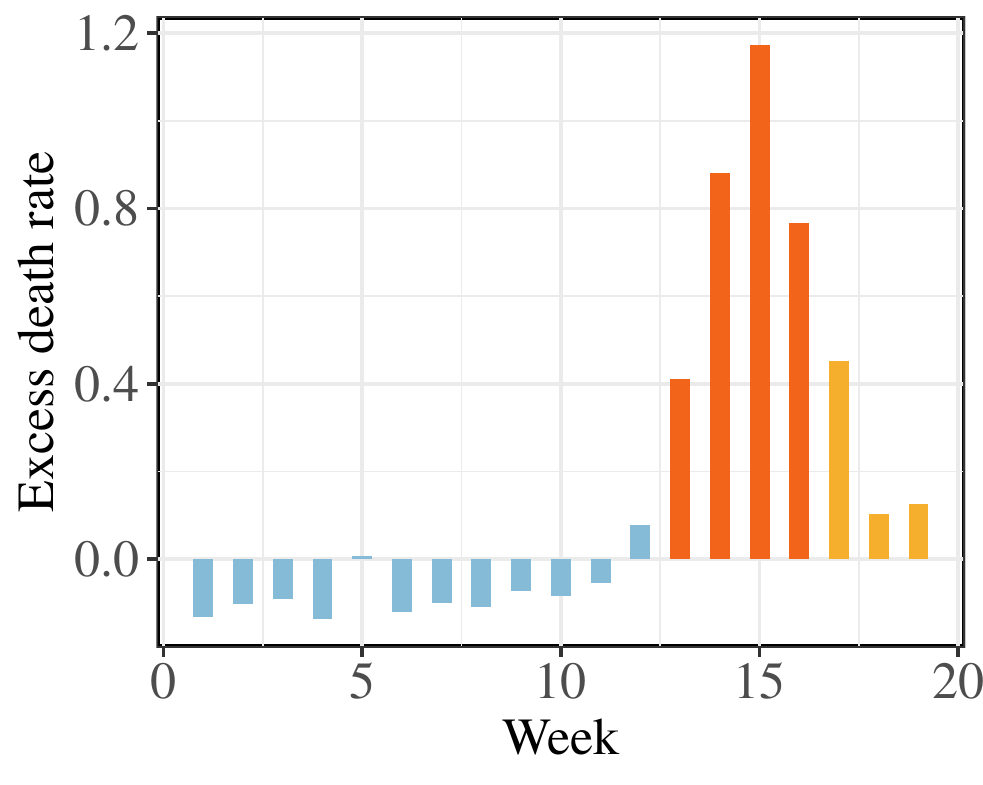}
                \label{p4}
    \end{subfigure}%
    \begin{subfigure}[b]{0.33\textwidth}
                \caption{Age over 85 (include 85)}
                \includegraphics[width=\linewidth]{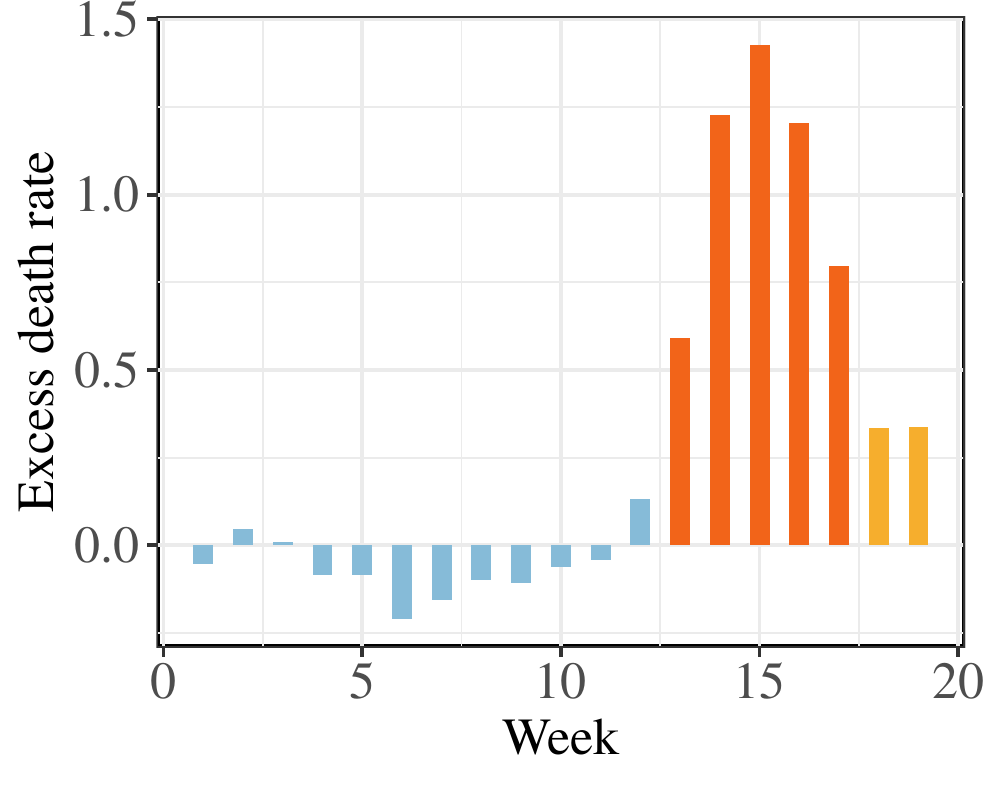}
                \label{p5}
    \end{subfigure}%
    \begin{subfigure}[b]{0.33\textwidth}
                \caption{All age group total}
                \includegraphics[width=\linewidth]{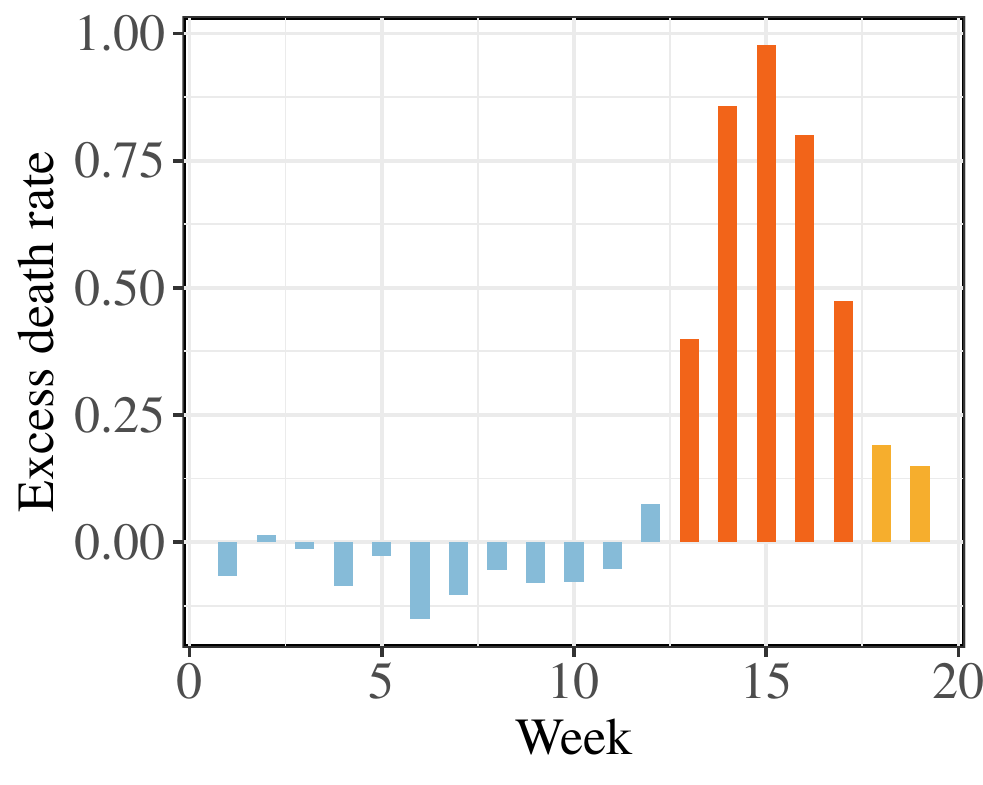}
                \label{p6}
    \end{subfigure}
    \caption{The weekly excess death rate in different age groups in 2020, with clusters labeled by colors}
\end{figure}

We first apply the change point detection to the matrix-valued time series $\{\mathbf{X}_t,t=1,\dots, T\}$. In this way, we obtain identical change points for all age groups and genders. There is no change point detected in the latter half of 2019. The change points detected in 2020 are in week 13 and week 18 (corresponds to $\widehat{t}^* = 39$ and $44$). Then, the excess death rate in 2020 fall into three clusters (see Figure~\ref{p6}). Week 12 is the end of the 'pre-COVID-19' period, and starting from week 13, the surge of weekly excess rates reveals the outbreak of COVID-19 among the population. From week 18, the weekly excess rate drops to one-fourth of the amount in the peak week (week 15). However, it is still too early to tell if the excess death curve is about to be flat since the collected data at the time of this study are only available up to week 19 of 2020.

As different age groups have different risk exposures to COVID-19, the change points might also vary across age groups. Therefore, we perform the detection for each age group separately, and the clusters are partitioned accordingly (see Figure~\ref{p1} - \ref{p5}).

We can see clear distinctions in the excess deaths distributions and the phases change points among the young population (age below 14), middle-aged population (age between 15 and 64), and the old population (age above 65). The young population's excess death rates are small, with negative numbers from week 5 (when the first COVID-19 case appeared in Belgium). One possible reason is that children have engaged in fewer outdoor activities and undertaken less international travel, making them less likely to contract the virus. Moreover, lock-down and social distancing policies mean they are less likely to be involved in accidents \citep[see also][]{lee2020children}. For people between 15 and 64, there is a steep increase in excess death rate in weeks 14 and 15; soon after that, the excess deaths drop to around zero. However, for older people, the COVID-19 deaths surge started one week earlier and lasted longer than the middle-aged population. This might be because older people are more vulnerable to COVID-19 due to their weak immunity and prior/on-going complications/diseases. Moreover, the older the age group is, the more excess death counts appeared, and the longer the ``COIVD-19 outbreak'' period lasted. Additionally, as over $60\%$ of the excess deaths are amongst those over age 85, the detected change points in the whole population are the same as those detected using the population above (and including) 85 only. 

\section{Conclusion} \label{s5}

We detect the COVID-19 pandemic phase in Belgium using a statistical change point detection technique. According to our analysis, the phase change shows age differentials. A surge in deaths in the elderly population appeared earlier than for the middle-aged population and also lasted longer. Policymakers should consider this evidence and adjust their policies and focus guidance towards specific age groups. For example, lock-down policies may be first relaxed for younger age groups, while older people should remain extremely cautious.

\bibliographystyle{agsm}
\bibliography{COVID19_mortality.bib}

\end{document}